\documentclass[showpacs,twocolumn,prl,superscriptaddress]{revtex4}
\usepackage{amsfonts}
\usepackage{amsmath}
\usepackage{amssymb}
\usepackage{graphicx}
\usepackage{verbatim}
\usepackage{textcomp}

\providecommand{\U}[1]{\protect\rule{.1in}{.1in}}

\begin{document}
\title{Quantum enhanced optomechanical magnetometry}

\author{Bei-Bei Li}
\affiliation{School of Mathematics and Physics, The University of Queensland, St Lucia, Queensland 4072, Australia.}
\affiliation{Qian Xuesen Laboratory of Space Technology,
Beijing 100094, P. R. China.}
\author{Jan Bilek}
\affiliation{Department of Physics, Technical University of Denmark, Fysikvej 309, 2800 Kgs. Lyngby, Denmark.}
\author{Ulrich B. Hoff}
\affiliation{Department of Physics, Technical University of Denmark, Fysikvej 309, 2800 Kgs. Lyngby, Denmark.}
\author{Lars S. Madsen}
\affiliation{School of Mathematics and Physics, The University of Queensland, St Lucia, Queensland 4072, Australia.}
\author{Stefan Forstner}
\affiliation{School of Mathematics and Physics, The University of Queensland, St Lucia, Queensland 4072, Australia.}
\author{Varun Prakash}
\affiliation{School of Mathematics and Physics, The University of Queensland, St Lucia, Queensland 4072, Australia.}
\author{Clemens Sch\"{a}fermeier}
\affiliation{Department of Physics, Technical University of Denmark, Fysikvej 309, 2800 Kgs. Lyngby, Denmark.}
\author{Tobias Gehring}
\affiliation{Department of Physics, Technical University of Denmark, Fysikvej 309, 2800 Kgs. Lyngby, Denmark.}
\author{Warwick P. Bowen}
\email{w.bowen@uq.edu.au}
\affiliation{School of Mathematics and Physics, The University of Queensland, St Lucia, Queensland 4072, Australia.}
\author{Ulrik L. Andersen}
\email{ulrik.andersen@fysik.dtu.dk}
\affiliation{Department of Physics, Technical University of Denmark, Fysikvej 309, 2800 Kgs. Lyngby, Denmark.}

\date{\today}

\begin{abstract}

The resonant enhancement of both mechanical and optical response in microcavity optomechanical devices allows exquisitely sensitive measurements of stimuli such as acceleration, mass and magnetic fields. In this work, we show that quantum correlated light can improve the performance of such sensors, increasing both their sensitivity and their bandwidth. Specifically, we develop a silicon-chip based cavity optomechanical magnetometer that incorporates phase squeezed light to suppress optical shot noise. At frequencies where shot noise is the dominant noise source this allows a 20\% improvement in magnetic field sensitivity. Furthermore, squeezed light broadens the range of frequencies at which thermal noise dominates, which has the effect of increasing the overall sensor bandwidth by 50\%. These proof-of-principle results open the door to apply quantum correlated light more broadly in chip-scale sensors and devices.



\end{abstract}



\maketitle
Cavity optomechanics \cite{2008Science,2014RMP,Warwick book} has attracted increasing research interest for both fundamental studies and practical applications. Strong radiation pressure coupling between high quality mechanical and optical resonances has enabled the demonstration of a range of interesting quantum behaviours, such as ground state cooling of macroscopic mechanical oscillators \cite{cooling1,cooling2,cooling3,cooling4}, quantum squeezing of mechanical motion \cite{SqeezMotion1,SqeezMotion2,SqeezMotion3,SqeezMotion4}, and the production of squeezed light \cite{SquezLightProd1,SquezLightProd2}; While the combination of resonance-enhanced mechanical and optical response \cite{2008NJP} has enabled precision sensors \cite{2014APR} ranging from kilometer-sized laser interferometer gravitational wave detectors \cite{LIGO,GW}, to micro/nano scale silicon chip based force \cite{force}, mass \cite{mass}, acceleration \cite{acceleration1,acceleration2}, and magnetic field \cite{magnetic1,magnetic2,magnetic3,magnetic4} sensors.

The precision of cavity optomechanical sensors is generally constrained by three fundamental noise sources: thermal noise from the environment, shot noise from the photon number fluctuations of the light used to probe the system, and quantum backaction noise arising from the radiation pressure of the probe light. With increasing laser power, the shot noise contribution decreases and backaction increases. An optimal sensitivity is reached when they are equal, termed the standard quantum limit (SQL) \cite{2008Science}. The noise floor can be engineered using quantum correlated light. For instance, squeezed light \cite{squeezing1,squeezing2,30years} allows the shot noise to be suppressed \cite{Caves}, thereby improving the sensitivity if the shot noise is dominant. Squeezed light has been used, for example, to improve the precision of gravitational wave interferometry in both LIGO and GEO \cite{GW1,GW2,GW3}, of nanoscale measurements of biological systems \cite{biology}, and of magnetic field measurement using atomic magnetometers \cite{atomic1,atomic2}. In cavity optomechanics, it has been used to enhance measurements of thermal noise \cite{toroid}, to improve both feedback \cite{feedback cooling} and sideband cooling \cite{sideband cooling}, and to study the backaction from the radiation pressure force \cite{strong radiation}. However, it has not previously been used to improve cavity optomechanical sensors of external stimuli. Here, we demonstrate the first application of squeezed light in such a sensor, specifically, in a cavity optomechanical magnetometer \cite{magnetic1,magnetic2}. At frequencies where shot noise is dominant, squeezed light suppresses the noise floor, improving the magnetic field sensitivity. Moreover, by increasing the range of frequencies over which thermal noise is dominant, the sensor bandwidth is also increased. Squeezed light enhanced sensor bandwidth \cite{bandwidth} is of importance in applications which need good sensitivity in a broadband range, e.g., in magnetic resonance imaging.\\

\begin{figure*}[ptb]
\begin{center}
\includegraphics[width=14cm]{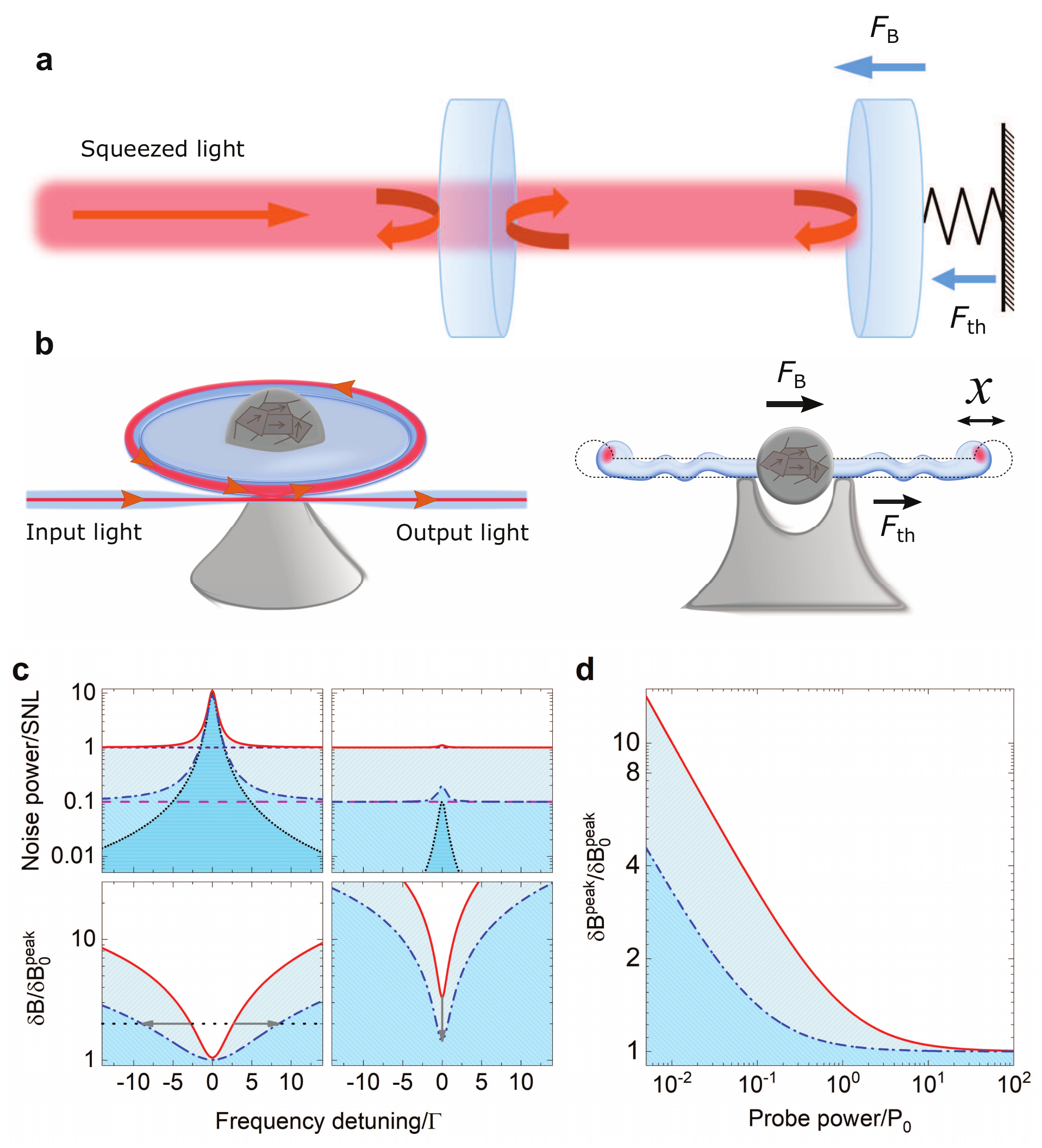}
\end{center}
\caption{\textbf{Conceptual schematic and theoretical results.} (\textbf{a}), Conceptual schematic of a cavity optomechanical system probed with squeezed light. Here $F_\mathrm{B}$ and $F_\mathrm{th}$ denote the magnetic field induced force and the thermal force on the mechanical oscillator. (\textbf{b}), Left: A schematic of a microtoroid magnetometer coupled with a nanofiber. Right: The cross section of a microtoroid, of which the optical field is distributed along the inner surface. The mechanical motion changes the circumference of the cavity, and thus shifts the optical resonance. (\textbf{c}-\textbf{d}), Theoretical result for squeezing enhanced performance of the magnetometer. Here we use a squeezing factor of 10~dB. In (\textbf{c}), top-left and bottom-left plots correspond to the strong probe power case, where $P=10P_\mathrm{0}$, with $P_{\mathrm{0}}$ defined as the power when the thermal noise on mechanical resonance equals the shot noise level (SNL), i.e., $\bar{n}=1/(16\eta C)$; while top-right and bottom-right plots correspond to the weak probe power case, where $P=0.1P_\mathrm{0}$. (\textbf{c}), Top left and top right, The noise power spectrum normalized to the SNL. Black short-dotted curve: thermal noise, purple short-dashed line: vacuum shot noise for coherent probe, magenta dashed line: squeezed vacuum noise for squeezed probe, red solid curve: total noise for coherent probe, blue dash-dotted curve: total noise for squeezed probe. Bottom left and Bottom right, The sensitivity as a function of frequency for coherent (red solid curves) and squeezed (blue dash-dotted curves) probe, respectively, normalized to $\delta B_\mathrm{0}^\mathrm{peak}$ which is the peak sensitivity for squeezed probe in the strong probe power case. (\textbf{d}), The peak sensitivity $\delta B^\mathrm{peak}$ (normalized to $\delta B_\mathrm{0}^\mathrm{peak}$) as a function of the probe power $P$, for coherent (red solid curve) and squeezed (blue dash-dotted curve) probes, respectively.}
\label{fig1}
\end{figure*}

\noindent {\bf Theoretical analysis}





Figure~\ref{fig1}a shows a conceptual schematic of a cavity optomechanical magnetometer, comprised of an optical cavity, coupled to a mechanical oscillator. The mechanical oscillator is driven by a force $F_{\mathrm{B}}$ induced by a magnetic field via the magnetostrictive effect \cite{magnetic1}, along with thermal and backaction noise forces. The mechanical motion of the oscillator changes the cavity length and thus the optical resonance. This modulates the phase of an injected squeezed probe field, and can therefore be read out via an optical phase measurement. In our case, the optical cavity is a microtoroid, whose circumference is modified by mechanical motion, as illustrated in Fig.~\ref{fig1}b. Our experiments operate in the unresolved sideband regime where the optical decay rate $\kappa$ is much larger than the mechanical resonance frequency $\Omega$. In this regime, the thermal force noise dominates the backaction noise when $\bar{n}>C$ \cite{Warwick book}, where $\bar{n}$ is the thermal phonon occupancy of the mechanical oscillator, and $C$ is the optomechanical cooperativity, which quantifies the strength of radiation pressure optomechanical coupling relative to the mechanical and optical dissipation rates and is proportional to the probe laser power. For the few megahertz frequencies we use, $\bar{n}\sim10^6$ at room temperature; while with the optical and mechanical properties of our optomechanical microresonator, and for the maximum optical power we use, $C\sim1000$. Consequently, the mechanical force noise is dominated by thermal noise, and we neglect backaction noise henceforth.

The displacement $x$ of the mechanical oscillator in response to an external force $F$ is quantified in the frequency domain by the mechanical susceptibility $\chi(\omega)$. To illustrate the physics, we consider the simple case of a single mechanical resonance, for which $\chi(\omega)=1/(m_{\mathrm{eff}}(\Omega^2-\omega^2-\mathrm{i}\omega\Gamma))$, where $m_{\mathrm{eff}}$ is the effective mass of the mechanical oscillator, and $\Gamma$ is its damping rate, enhancing the mechanical response to near resonant forces (see top left and top right of Fig.~\ref{fig1}c). Quite generally in cavity optomechanical sensors, away from resonance, optical shot noise is dominant, allowing squeezed-light enhanced sensitivity; while for a single-sided cavity in the unresolved sideband regime, thermal noise dominates shot noise at resonance if $\bar{n}>1/(16\eta C)$, where $\eta$ is the optical detection efficiency.

A magnetic field is resolvable when the signal it induces is larger than the total noise floor. Neglecting backaction noise, this leads to a minimum detectable force $\delta F$
\begin{equation}\label{1}
\delta F=\sqrt{2m_{\mathrm{eff}}\Gamma k_{\mathrm{B}}T}\left[1+\frac{V_{\mathrm{sqz}}}{16\bar{n}C}\left|\frac{\chi(\Omega)}{\chi(\omega)}\right|^2\right]^{1/2}
\end{equation}
\noindent for a cavity without internal losses and with a perfect optical detection efficiency $\eta=1$. $k_{\mathrm{B}}$ and $T$ are the Boltzmann constant and the temperature, respectively. The first term in the bracket on the right hand side represents the thermal noise, while the second term represents the optical noise, with $V_{\mathrm{sqz}}$ the squeezed quadrature variance of the squeezed light. Introducing an actuation constant $c_{\mathrm{act}}=F/B$ which characterizes how well the magnetic field $B$ is converted into an applied force $F$ on the mechanical oscillator \cite{magnetic1}, the magnetic field sensitivity is $\delta B=\delta F/c_{\mathrm{act}}$.

From Eq.~\ref{1} we see that the peak sensitivity occurs on mechanical resonance. In the case where thermal noise is dominant at mechanical resonance frequency ($\bar{n}>1/16C$), squeezed light does not significantly change the peak sensitivity, instead extending the frequency range over which thermal noise dominates, and therefore the sensor bandwidth (bottom left of Fig.~\ref{fig1}c); while in the case where optical noise is dominant on resonance ($\bar{n}<1/16C$), both the peak sensitivity and bandwidth are improved by squeezed light (bottom right of Fig.~\ref{fig1}c). The saturation of sensitivity to the optimal (thermal noise limited) sensitivity as probe powers increase is shown in Fig.~\ref{fig1}d. It can be seen that squeezed light reduces the probe power required to reach the optimal sensitivity.\\

\noindent {\bf Results}

\noindent {\bf Measurement of the optomechanical system.} The optomechanical magnetometer is a microtoroid cavity with a grain of magnetostrictive material (terfenol-D) \cite{magnetic1,magnetic2}, as sketched in Fig.\ref{fig1}b. In such magnetometers, the magnetic field deforms the microcavity via the magnetostrictive expansion and shifts the optical resonance. In the case of an alternating current (AC) magnetic field, the magnetostrictive material exerts a periodic force on the mechanical oscillator, which can drive the mechanical motion of the toroid. When the microcavity is excited on optical resonance, the mechanical motion translates into a pure phase modulation of the transmitted light at the mechanical frequency, which is read out with a homodyne detector and recorded using a spectrum analyzer.

\begin{figure}[ptb]
\begin{center}
\includegraphics[width=8.5cm]{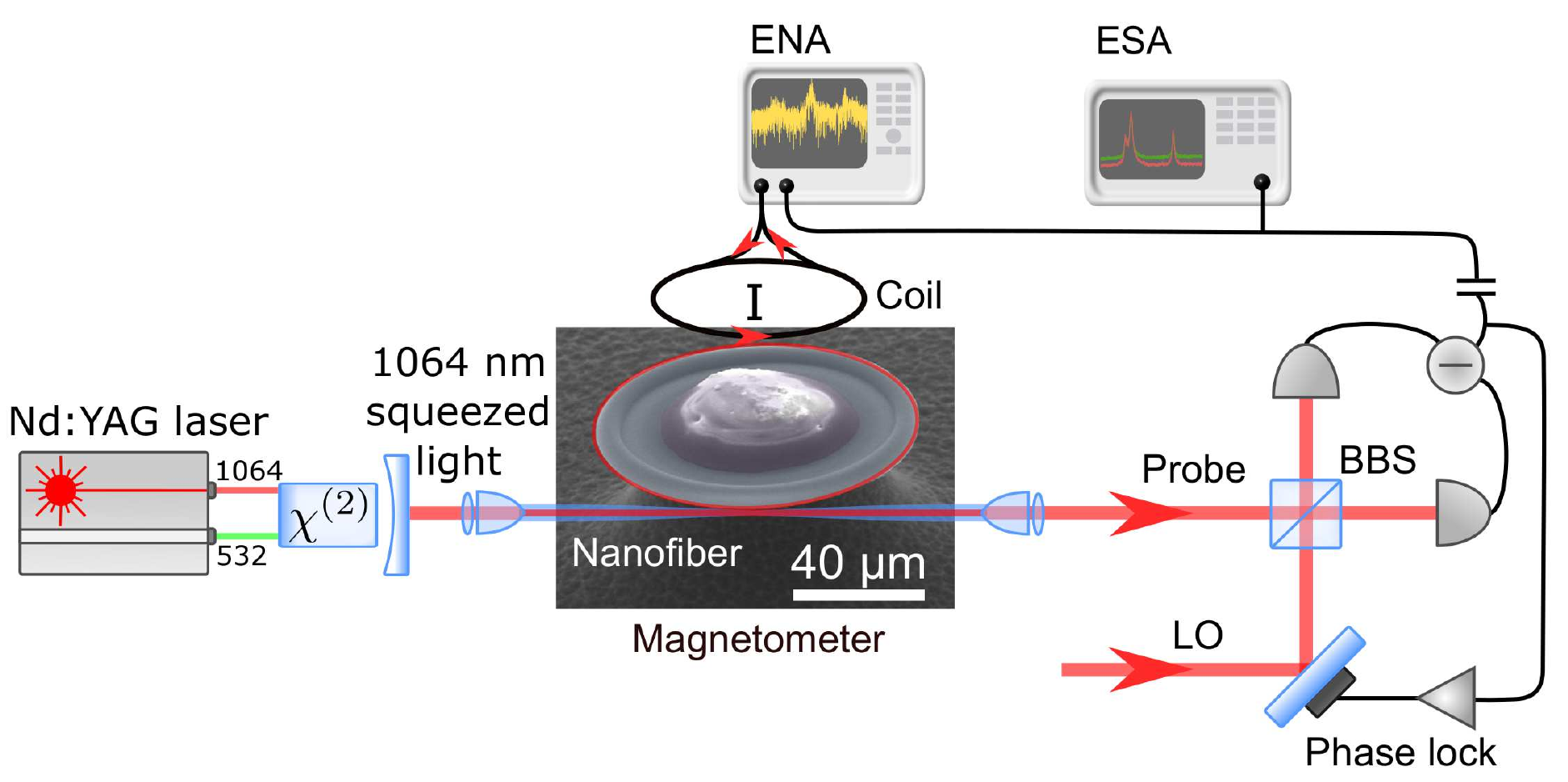}
\end{center}
\caption{\textbf{Measurement setup for squeezed light enhanced cavity optomechanical magnetometry.} Squeezed light at a wavelength of 1064~nm is used to probe the magnetometer (see Methods). The magnetometer is a microtoroid with terfenol-D embedded inside, as shown in the scanning electron microscope picture. The optical Q factor of the toroid mode is about $1\times10^6$, corresponding to an optical damping rate of $\kappa/2\pi\sim300$~MHz. The mechanical motion of the toroid is measured by performing a homodyne detection. LO: local oscillator, BBS: balanced beam splitter, comprised of two polarization beam splitters and a half wave plate, ESA: electronic spectrum analyzer, ENA: electronic network analyzer.}
\label{fig2}
\end{figure}

The measurement setup for squeezed light enhanced magnetometry is shown in Fig.~\ref{fig2}. A Nd:YAG laser is used to produce squeezed light at a wavelength of 1064~nm. The light is coupled into the microtoroid evanescently through an optical nanofiber with a diameter of about 700~nm. The optical resonance of the cavity is thermally tuned to match the wavelength of the laser. The cavity phase is actively locked using a feedback system \cite{PDH}. A coil is used to produce an AC magnetic field to test the magnetic field response of the magnetometer. The mechanical motion of the toroid is measured by performing homodyne detection (see Methods for more details). An electronic spectrum analyzer (ESA) is used to record the noise power spectrum. In order to measure the response of the magnetometer to magnetic fields at different frequencies, we drive the coil with the output of an electric network analyzer (ENA) and measure the magnetic field response at each frequency with the same ENA.\\
\\

\begin{figure}[ptb]
\begin{center}
\includegraphics[width=8cm]{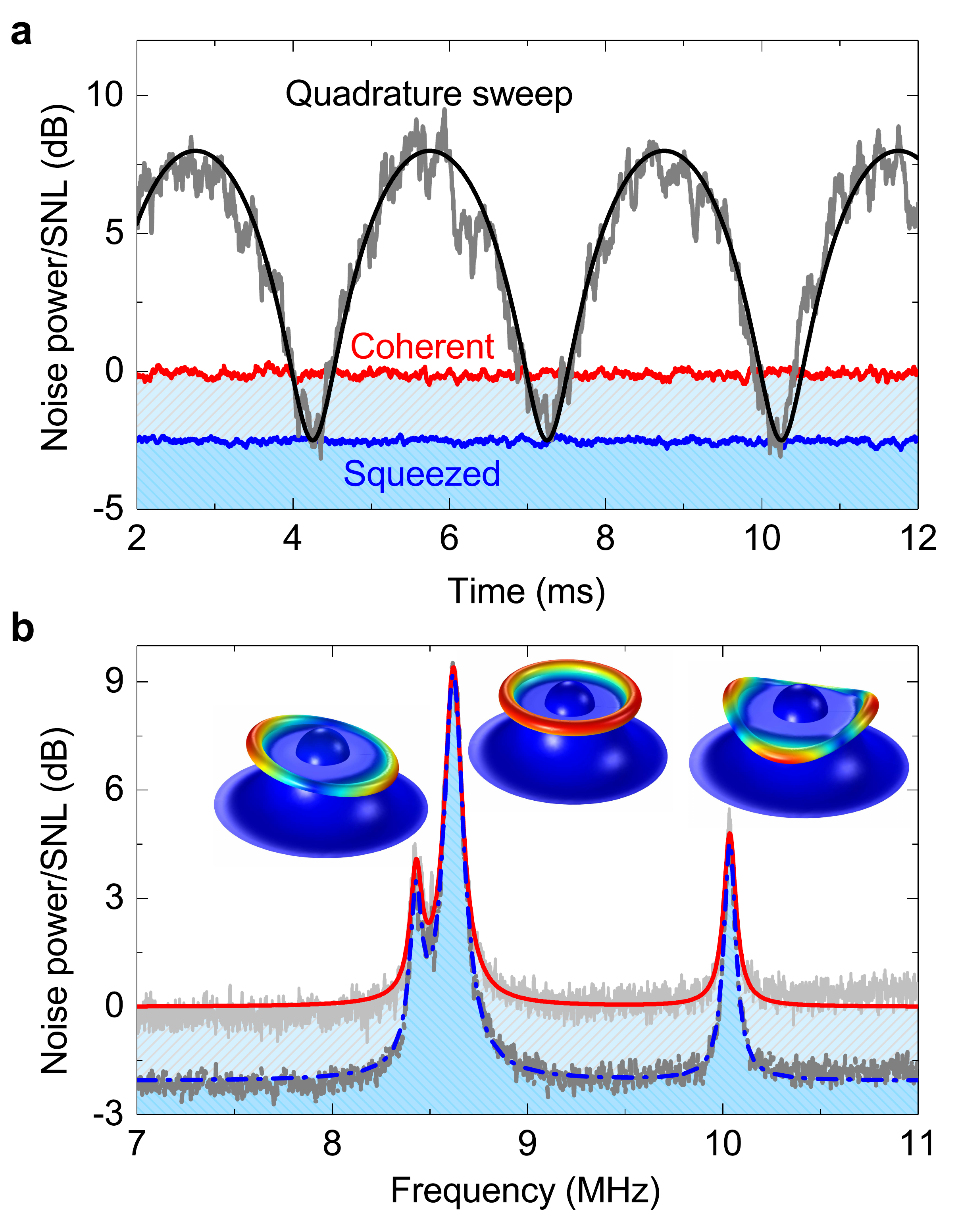}
\end{center}
\caption{\textbf{Characterization of the squeezed light.} (\textbf{a}), Characterization of the squeezed state before coupling to the microtoroid. The dark grey curve shows the noise power when sweeping LO phase continuously, with its theoretically fitted result shown in the black solid curve. The red and blue solid curves are the noise power with coherent and squeezed probe, when the LO phase is locked at the phase quadrature. (\textbf{b}), The measured noise power from the microtoroid, with both coherent (light grey curve) and squeezed (dark grey curve) probe, respectively. The red solid and the blue dashed curves are the fitted results for the measured ones. The three peaks correspond to three mechanical resonance modes (from left to right: tilting mode, flapping mode, and crown mode), with the profiles shown in the inset, obtained using COMSOL Multiphysics.}
\label{fig3}
\end{figure}

\noindent {\bf Characterization of the squeezed light.} To characterize the squeezed state transmitted through the fiber, we decouple the microtoroid from the nanofiber and measure the homodyne detection signal of the field quadratures by linearly sweeping the LO phase $\theta$. As shown in the dark grey curve in Fig.~\ref{fig3}a, when $\theta$ is swept continuously, the noise power changes periodically, following the equation $V=V_\mathrm{sqz}\mathrm{cos}^2\theta+V_\mathrm{anti}\mathrm{sin}^2\mathrm{\theta}$, with $V_\mathrm{anti}$ being the anti-squeezed quadrature variance. The black solid curve is the fitted result based on this equation, yielding $V_\mathrm{sqz}=0.56$, and $V_\mathrm{anti}=6.3$. Ideally, the product $V_{\mathrm{sqz}}V_{\mathrm{anti}}=1$, satisfying the Heisenberg uncertainty limit, but in reality this limit is not reached, due to loss of the squeezed light during propagation in the setup. The noise power reaches its minimum when locked at the phase quadrature, and we lock $\theta$ to that quadrature henceforth. The red and blue curves in Fig.~\ref{fig3}a show the noise power for phase quadrature measurement of coherent and squeezed probes, respectively.\\
\\
\noindent {\bf Magnetic field measurement with a coherent or squeezed probe.} The squeezed field is coupled into the microcavity through the nanofiber. We keep the fiber-cavity coupling in the undercoupled regime, in which case most of the squeezing is preserved. The noise power with both coherent and squeezed probes in the frequency range of 7-11~MHz is measured, as shown in the light grey (for coherent probe) and grey (for squeezed probe) curves in Fig.~\ref{fig3}b. With a probe power of 80~\textmu W, three peaks appear in this frequency range of the noise spectrum, corresponding to three thermally excited mechanical resonance modes. We use COMSOL Multiphysics simulations to identify these three modes as tilting mode, flapping mode, and crown mode, with the corresponding mode profiles shown in the inset. It can be seen that over the frequency ranges where the optical noise dominates, the noise floor is suppressed by up to 2.2~dB by squeezed light, while it is left essentially unchanged when thermal noise dominates.

\begin{figure}[ptb]
\begin{center}
\includegraphics[width=8cm]{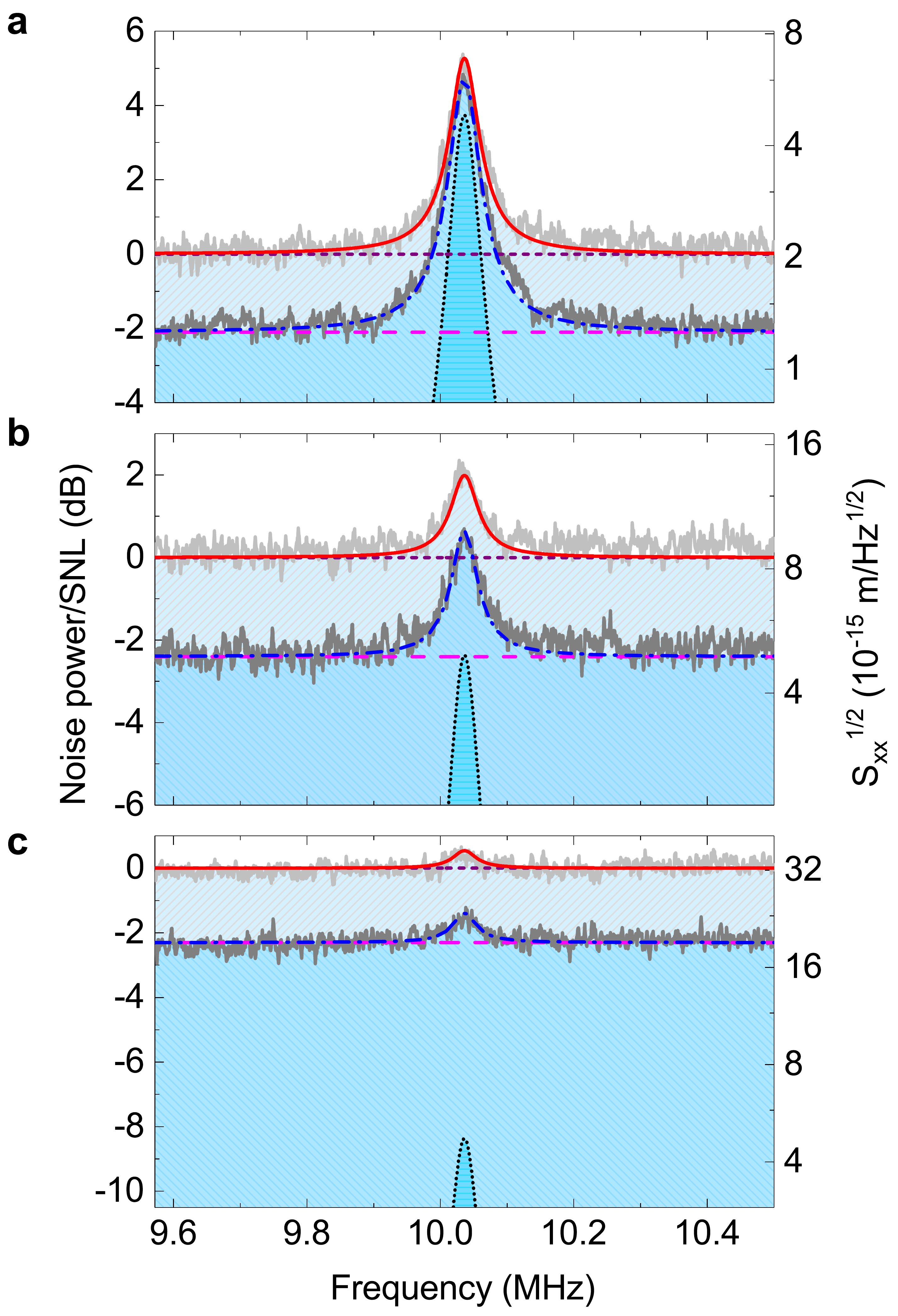}
\end{center}
\caption{\textbf{Noise power spectra measurement with coherent and squeezed probes.} Characterization of the noise power spectra around the crown mode, under different probe powers: (\textbf{a}) 80 \textmu W, (\textbf{b}) 20 \textmu W, and (\textbf{c}) 5 \textmu W. The light grey and dark grey curves are the measured noise power for coherent and squeezed probe, respectively. The other curves are the theoretically fitted ones: black short-dotted curves: thermal noise, purple short-dashed lines: vacuum shot noise with coherent probe, magenta dashed lines: squeezed vacuum noise with squeezed probe, red solid curves: total noise for the coherent probe, and the blue dash-dotted curves: total noise for the squeezed probe. On the right axes of the figures, it shows the corresponding displacement amplitude spectral density $S_{xx}^\mathrm{1/2}$. The mechanical damping rate is extracted from the linewidth of the mode in the thermal noise spectrum, to be $\Gamma/\mathrm{2\pi}=42$~kHz. The effective mass of the crown mode is determined to be $m_\mathrm{eff}=6.06$~ng obtained from COMSOL modeling. The displacement amplitude spectral density $S_{xx}^\mathrm{1/2}$ is plotted on the right axes of the figures.}
\label{fig4}
\end{figure}




In order to carefully study the effect of the probe power on the noise spectrum, in the following we focus on the crown mode with mechanical resonance frequency of $\Omega/2\pi=\mathrm{10.035}$ MHz. Figure~\ref{fig4}a shows the noise (normalized to the shot noise level) in the vicinity of the crown mode with probe power $P$=80~\textmu W. As expected, in this case the noise level remains unchanged by squeezing near the resonance frequency where thermal noise is dominant, and is suppressed away from resonance. As the probe power gradually decreases, the thermal noise drops relative to the shot noise. As shown in Figs.~\ref{fig4}b and c, the shot noise is dominant in the whole frequency range, for probe powers of 20~\textmu W and 5~\textmu W. At these power levels, squeezing allows the noise floor to be suppressed over the entire frequency ranges. These results are consistent with the predictions in Fig.~\ref{fig1}c.




\begin{figure}[ptb]
\begin{center}
\includegraphics[width=8cm]{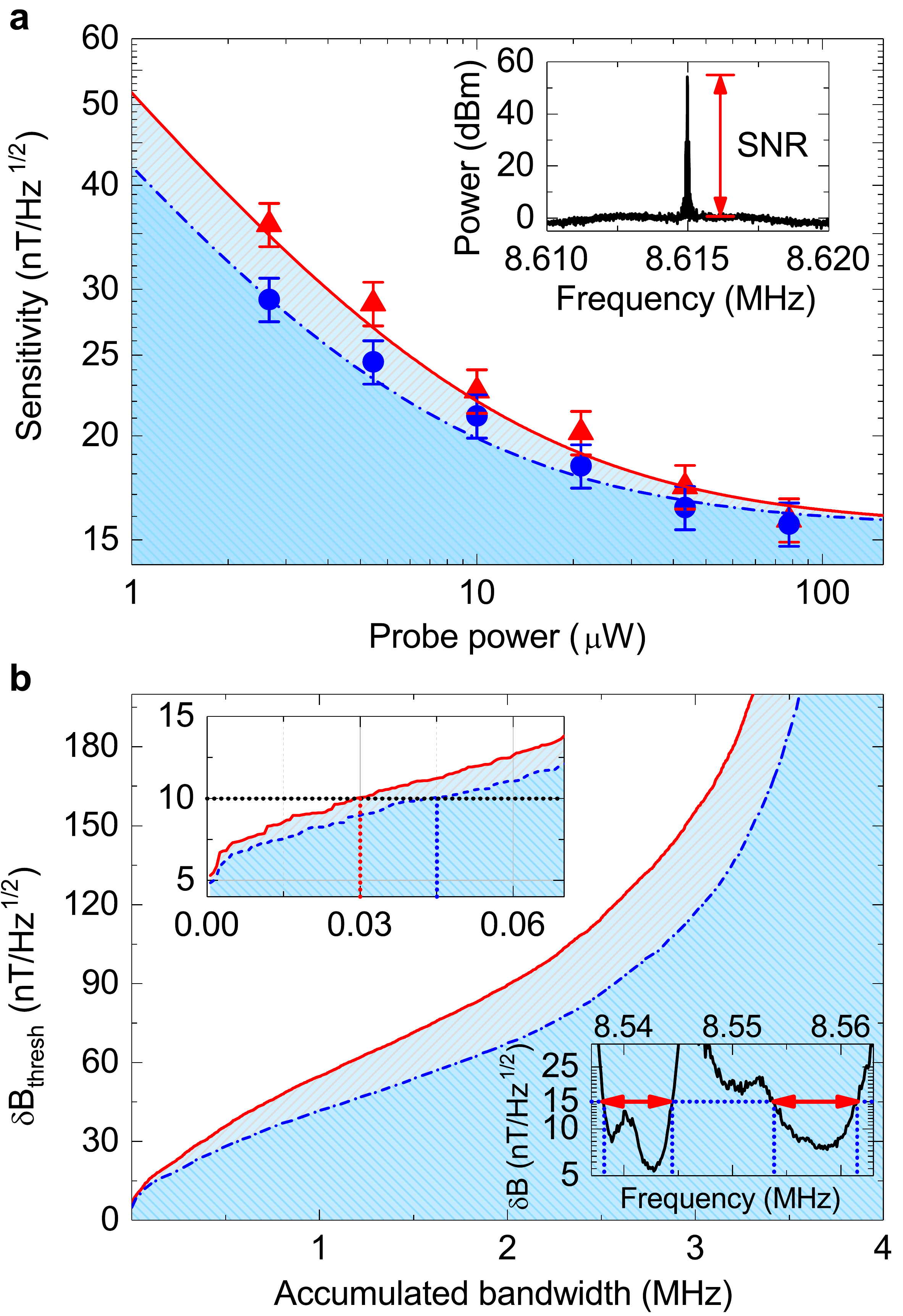}
\end{center}
\caption{\textbf{Sensitivity and bandwidth improvement.} (\textbf{a}), Sensitivity at the frequency of 8.615 MHz, as a function of the probe
power. The red triangles and blue circles represent the measured results for coherent and squeezed probes, respectively. The error bars are obtained by taking into account the fluctuation in the noise power measurement. The red solid (coherent) and blue dash-dotted (squeezed) curves are the corresponding theoretical fitted result. The inset shows the power spectrum when the magnetometer is driven at this frequency, with the peak denoting the signal induced by the magnetic field. (\textbf{b}), The accumulated bandwidth as a function of the threshold sensitivity, for the coherent (red solid curve) and squeezed (blue dashed curve) probe, respectively. Top-left inset: the zoom-in of the accumulated bandwidth in the frequency range of 0-0.07~MHz, showing the 3-dB bandwidth of 30~kHz for the coherent probe and 45~kHz for the squeezed probe. Bottom-right inset: The sensitivity spectrum in the frequency range of 8.537-8.563~MHz, showing the definition of the accumulated bandwidth. For a threshold sensitivity of 15~nT/$\sqrt{\mathrm{Hz}}$, the accumulated bandwidth is defined as the sum of the frequency ranges within the two red arrows.}
\label{fig5}
\end{figure}

The magnetic field sensitivity of the magnetometer is then characterized (see Methods for more details). We first characterize the absolute sensitivity at a single frequency $\omega_{\mathrm{ref}}$=8.615~MHz. The inset of Fig.~\ref{fig5}a shows the power spectrum at $\omega_{\mathrm{ref}}$, when the magnetometer is driven with a magnetic field with known strength $B_\mathrm{ref}$. The sensitivity at this frequency can be derived from the signal-to-noise ratio (SNR) and $B_\mathrm{ref}$, $\delta B_{\mathrm{ref}}=B_\mathrm{ref}/\sqrt{\mathrm{RBW}\times\mathrm{SNR}}$ \cite{magnetic1}, with RBW being the measurement resolution bandwidth. Figure~\ref{fig5}a plots the sensitivity at this frequency as a function of the probe power. The red triangles and the blue circles represent the measured result for coherent and squeezed probes, respectively, with the error bars obtained by taking into account the fluctuation of about $\pm0.5$~dB in the measured noise spectrum. As expected, the sensitivity is improved by squeezing at low probe power where the shot noise is dominant and reaches the same optimal sensitivity at high probe power where the thermal noise is dominant, in good agreement with theoretical fits. For instance, the sensitivity at 2.5~\textmu W probe power is improved from 35.9~nT/$\sqrt{\mathrm{Hz}}$ to 29.2~nT/$\sqrt{\mathrm{Hz}}$, and thermal noise limited sensitivity is about 15.7~nT/$\sqrt{\mathrm{Hz}}$ for both coherent and squeezed probes. The sensitivity at $\omega_{\mathrm{ref}}$ can be used along with network analysis to calibrate the sensitivity over the whole frequency range (see Methods). This allows the effect of squeezing on bandwidth to be analyzed, as discussed in the following. For a probe power of 80~\textmu W, the peak sensitivity in the whole frequency range is found to be about 5~nT/$\sqrt{\mathrm{Hz}}$ at $\omega/2\pi\sim$ 8.543~MHz for both coherent and squeezed probes, at the power of 80~\textmu W.

The sensitivity is found to vary significantly over frequency ranges of around 10~kHz, due to resonances in the response of terfenol-D, as shown in the sensitivity spectrum in the bottom-right inset of Fig.~\ref{fig5}b, and consistent with previous observations \cite{magnetic1}. This precludes comparison of the magnetometer bandwidth as a function of squeezing to a simple theory. Instead here we analyze the squeezing dependence of the accumulated bandwidth, defined as the total frequency range over which the sensitivity is better than a certain threshold value $\delta B_{\mathrm{thresh}}$ (see bottom-right inset). In Fig.~\ref{fig5}b, we plot the accumulated bandwidth for coherent (red solid curve) and squeezed (blue dash-dotted curve) probes, at a probe power of 80~\textmu W. It can be seen that, for each $\delta B_{\mathrm{thresh}}$, the accumulated bandwidth for the squeezed probe is greater than that for the coherent probe. The upper-left inset of Fig.~\ref{fig5}b shows the accumulated bandwidth over the smaller frequency range of 0-70~kHz. Squeezed light expands the 3~dB bandwidth (corresponding to $\delta B_\mathrm{thresh}=10~$nT/$\sqrt{\mathrm{Hz}}$) by 50\%, from 30~kHz (for coherent probe) to 45~kHz (for squeezed probe).\\
\\

\noindent {\bf Conclusions}

In summary, we have demonstrated the first application of quantum light in a microcavity optomechanical sensor. By probing a cavity optomechanical magnetometer with phase squeezed light, the noise floor is suppressed by about 40\%, allowing improved sensitivity by about 20\% in the shot noise dominated regime, and a 50\% enhancement in accumulated bandwidth from 30~kHz to 45~kHz. Squeezed light, further, reduces the optical power required to reach the optimal sensitivity.

Our approach provides a way to improve the sensitivity of the cavity optomechanical magnetometer over a broad frequency range, and also opens up possibilities for improving other optomechanical sensors, e.g., inertial sensors \cite{acceleration1,acceleration2}. While a 20\% improvement in sensitivity is relatively modest, recent advances in squeezing technologies \cite{squeezing6dB,squeezing9dB,squeezing12.7dB,squeezing15dB} hold promise for more substantial improvements. For instance, with squeezing of 15~dB recently reported \cite{squeezing15dB}, a sensitivity improvement of a factor of 5.6 could potentially be realized. Moreover, squeezed light could be generated on the same silicon chip as the sensor itself, using either radiation pressure induced optomechanical effects \cite{SquezLightProd1,SquezLightProd2} or nonlinear waveguides \cite{Nonlinear waveguide}. Further improvements may be possible by optimizing the magnetometer design itself, with sensitivities on the order of 100~pT/$\sqrt{\mathrm{Hz}}$ reported in previous cavity optomechanical magnetometers \cite{magnetic2}. Sensitivities in this range make cavity optomechanical magnetometers a promising candidate for a range of applications such as on-chip microfluidic nuclear magnetic resonance for medical diagnosis \cite{NMR} and magnetoencephalography \cite{MEG}, without the requirement for cryogenic systems, necessary for other precision magnetometers, such as superconducting quantum interference device (SQUID) based magnetometers \cite{SQUID1,SQUID2}.\\
\\

\noindent {\bf Methods}

\noindent {\bf Generation of squeezed light.}
Phase-squeezed light is generated through a parametric down conversion process in a 10mm PPKTP crystal enclosed in a linear cavity \cite{feedback cooling}. As shown in Fig. \ref{fig2}, both the 532~nm light (the pump light) and 1064~nm light (the seed light) are injected to the cavity. To generate phase squeezed light, the pump phase is locked to the seed beam amplification. \\
\\
\noindent {\bf Homodyne detection.} The balanced homodyne detector combines two inputs: a relatively weak probe which couples with the microcavity and a relatively strong local oscillator (LO) which comes from the same laser but without going through the microcavity. The homodyne detection signal is proportional to the product of the probe power $P$ and local oscillator power $P_{\rm{LO}}$. In our experiment, we keep $P_\mathrm{LO}=5$~mW, and vary the probe power $P$ from 1 to 100~\textmu W.\\
\\
\noindent {\bf Measurement of the magnetic field sensitivity.} The magnetometer is fabricated by embedding a grain of magnetostrictive material (terfenol-D) into the microtoroid \cite{magnetic2}. In order to obtain the sensitivity over the whole frequency range, we first measure the sensitivity at one reference frequency $S_\mathrm{ref}$, and use it to calibrate the sensitivity over the whole spectrum $S_\omega$. The inset of Fig.~\ref{fig5}a shows the power spectrum when the magnetometer is driven with a magnetic field with known strength $B_\mathrm{ref}$ at the reference frequency 8.615~MHz, from which the sensitivity at this frequency $S_\mathrm{ref}$ is derived. We then sweep the frequency of the magnetic field and measure the response in the whole frequency range, $R_{\omega}$. Then the sensitivity over the whole frequency range, $S_{\omega}$, is derived from $S_\mathrm{ref}$, $R_{\omega}$ and $N_{\omega}$ (the noise spectrum), $S_{\omega}=S_\mathrm{ref}\sqrt{(N_\omega R_\mathrm{ref})/(N_\mathrm{ref}R_{\omega})}$. The magnetic field signal at each frequency is the same for coherent and squeezed probes, as it only depends on the probe power and the properties associated with the magnetometer itself (including the magnetostrictive coefficient, coupling between the motion of the terfenol-D and the toroid, mechanical quality factor, optical quality factor, and the optomechanical coupling strength). Therefore, the sensitivity, which is inversely proportional to SNR, gets improved for a lower noise level.\\
\\

\bigskip

\noindent {\bf Acknowledgements}\\
We thank James Bennett, Rachpon Kalra, Christopher Baker, Andreas Sawadsky, Alexander Huck, Jonas Neergaard-Nielsen, and Halina Rubinsztein-Dunlop for the very helpful discussions. We thank the support from the Villum Foundation (Grant No. 13300), the Danish National Research Foundation (bigQ), DARPA QuASAR Program, Australian Research Council Discovery Project DP140100734, and Australian Defence Science and Technology Group projects CERA49 and CERA50. W.P.B. acknowledges the Australian Research Future Fellowship FT140100650. B.B.L. also acknowledges the support from the University of Queensland Postdoctoral Research Fellowship (609322) and Natural Science Foundation of China (NSFC 61705259 and 11654003). T.G. acknowledges support from the Danish Research Council for Independent Research (Individual Postdoc and Sapere Aude, 4184-00338B). Device fabrication was performed within the Queensland Node of the Australian Nanofabrication Facility.\\
Bei-Bei Li and Jan Bilek contributed equally to this work.\\
\\
\noindent {\bf Author contributions}\\
U.L.A. and W.P.B. conceived the idea while B.B.L., J.B., U.B.H.,
L.S.M., T.G., W.P.B., U.L.A. devised the experiment; B.B.L. and J.B. performed the experiment with the help from U.B.H, T.G., L.S.M., C.S., S.F. and V.P.; B.B.L. and J.B. analyzed the data with help from W.P.B.; B.B.L. and J.B. wrote the manuscript with help from W.P.B.; All authors contributed to the discussion and provided useful feedback on the paper; W.P.B. led the device fabrication part and U.L.A. led the optical measurement part; T.G., W.P.B. and U.L.A. supervised the whole work.\\
\\
\noindent {\bf Competing financial interests:} The authors declare no competing financial interests.

\end{document}